\documentclass[acmsmall,screen]{acmart}
\usepackage{hyperref}
\usepackage{multirow}
\usepackage{amsmath}
\usepackage[capitalise]{cleveref}
\usepackage{textcomp}
\usepackage{xcolor}
\usepackage{graphicx}
\usepackage{booktabs}
\usepackage{listings}
\usepackage[most]{tcolorbox}
\usepackage{enumitem}
\usepackage{makecell}
\usepackage[normalem]{ulem}
\usepackage{algorithm}
\usepackage{algpseudocode}

\definecolor{promptbg}{RGB}{248,249,251}
\definecolor{promptframe}{RGB}{95,99,104}
\definecolor{prompttitle}{RGB}{36,92,160}
\definecolor{promptkeyword}{RGB}{128,0,128}
\definecolor{promptcomment}{RGB}{90,110,90}

\lstdefinestyle{promptstyle}{
  basicstyle=\ttfamily\footnotesize,
  breaklines=true,
  breakatwhitespace=false,
  columns=fullflexible,
  keepspaces=true,
  showstringspaces=false,
  tabsize=2,
  frame=none,
  numbers=none,
  xleftmargin=0pt,
  xrightmargin=0pt,
  aboveskip=0pt,
  belowskip=0pt,
  keywordstyle=\color{promptkeyword}\bfseries,
  commentstyle=\color{promptcomment}\itshape,
  literate=
    {->}{{$\rightarrow$}}2
    {>=}{{$\geq$}}2
    {<=}{{$\leq$}}2
}

% ...
\newtcolorbox{promptbox}[2]{
  enhanced,
  breakable,
  colback=promptbg,
  colframe=promptframe,
  boxrule=0.5pt,
  arc=1.5mm,
  left=1.5mm,
  right=1.5mm,
  top=1.2mm,
  bottom=1.2mm,
  before skip=5pt,
  after skip=5pt,
  title={\textbf{#1}},
  fonttitle=\small\bfseries,
  coltitle=prompttitle,
  attach boxed title to top left={
    xshift=2mm,
    yshift=-1.5mm
  },
  boxed title style={
    colback=white,
    colframe=white,
    boxrule=0pt,
    left=0.5mm,
    right=0.5mm,
    top=0mm,
    bottom=0mm
  },
  label={#2}
}

\def\BibTeX{{\rm B\kern-.05em{\sc i\kern-.025em b}\kern-.08em
    T\kern-.1667em\lower.7ex\hbox{E}\kern-.125emX}}
\renewcommand{\arraystretch}{1.3}

\newcommand{\improve}[1]{\textcolor{black}{#1}}
\robustify{\improve}
\newcommand{\tool}{DesignEvo}

\usepackage[most]{tcolorbox}
\newtcolorbox{myquote}{
  colback=gray!20,   
  colframe=black!75!white,  
  fonttitle=\bfseries,       
  left=1em, right=1em,       
  boxrule=0pt,            
  arc=8pt,                   
  auto outer arc,            
  lifted shadow={1mm}{-1mm}{2mm}{0.3}  
}

\usepackage{subcaption}

\begin{document}

\title{DesignEvo: Recovering and Tracking Repository-Level Implemented Design Systems Across Commits}

\author{Mingyue Yuan}
\affiliation{%
  \institution{The University of New South Wales}
  \country{Australia}
}
\email{mingyue.yuan@unsw.edu.au}

\author{Jinbo Yang}
\affiliation{%
  \institution{Beijing Institute of Technology}
  \country{China}
}
\email{jinboy@bit.edu.cn}

\author{Jieshan Chen}
\affiliation{%
  \institution{CSIRO}
  \country{Australia}
}
\email{jieshan.chen@csiro.au}

\author{Aorui Yang}
\affiliation{%
  \institution{Zhejiang University}
  \country{China}
}
\email{22551167@zju.edu.cn}

\author{Aaron Quigley}
\affiliation{%
  \institution{The Australian National University}
  \country{Australia}
}
\email{aaron.quigley@anu.edu.au}

\author{Gelareh Mohammadi}
\affiliation{%
  \institution{The University of New South Wales}
  \country{Australia}
}
\email{g.mohammadi@unsw.edu.au}

\author{Zhenchang Xing}
\affiliation{%
  \institution{CSIRO}
  \country{Australia}
}
\email{zhenchang.xing@data61.csiro.au}

\author{Boxi Wu}
\affiliation{%
  \institution{Zhejiang University}
  \country{China}
}
\email{wuboxi@zju.edu.cn}

\author{Dehai Zhao}
\authornote{Corresponding author.}
\affiliation{%
  \institution{Zhejiang University}
  \country{China}
}
\email{dehai.zhao@zju.edu.cn}

\begin{abstract}
\improve{Modern frontend applications rely on design systems to maintain consistency across pages and releases. However, documentation and component libraries do not fully capture how interface patterns are implemented, reused, and adapted in practice. Existing UI analysis techniques focus on visual defects, interface structure, or element correspondences, providing limited support for automatically recovering and tracking implemented design systems at the repository level. To bridge this gap, we present DesignEvo, which combines screenshots, DOM structure, element geometry, and style values to recover functional UI groups. DesignEvo constrains vision-language model judgments to program-generated candidate regions, aggregates recovered groups and styles into repository-level records, and infers lifecycle states across adjacent commits. We evaluate DesignEvo on 10 open-source repositories comprising 317 page-version instances, 3,478 annotated functional UI groups, and 2,435 lifecycle-state labels. DesignEvo achieves an F1 score of 0.85 for group recovery and an overall lifecycle-state accuracy of 0.84, compared with 0.71 for the strongest lifecycle baseline. Ablations demonstrate the contributions of candidate constraints, functional categories, and style evidence. The resulting records provide structured, traceable evidence of recurring UI patterns and their evolution to support frontend maintenance and change auditing.}
\end{abstract}


\maketitle

\section{Introduction}

Modern software user interfaces (UIs) are normally built on top of internal \textbf{design systems}: reusable component libraries, style values, and interaction conventions that encode the visual, structural, and behavioral rules the product's UI is expected to follow~\cite{material_design}. From a software engineering standpoint, a design system serves as the UI layer's \textbf{quality contract}: an evolving specification that hundreds of pages and dozens of engineers are jointly expected to conform to. Its presence shapes downstream UI quality assurance, accessibility checking, cross-team governance, and long-term frontend maintenance~\cite{yuan2025designrepair,chiou2024reflow,mesbah2012automated,mazinanian2014discovering,ferreira2024refactoring,zeroheight_2026_report}. However, documented conventions do not always reflect how interface patterns are implemented and adapted in practice. As frontend applications evolve, locally modified components, inconsistent styling, and outdated patterns can accumulate across pages, making it increasingly difficult to understand and maintain UI consistency.

A design system is only \textbf{partially} documented. Its authoritative form lives in the frontend implementation—the components, class hierarchies, style tokens, and layout patterns actually rendered to users. The documented design system and the \textit{implemented} design system are two artifacts that must be kept aligned, and in practice they routinely diverge. Engineers introduce a locally reasonable Card variant that never returns to the shared library; a color is hard-coded instead of referencing the shared token; a deprecated pattern lingers in three pages after the design system has retired it; a de facto convention emerges in code that the documentation has not yet acknowledged. We refer to all such gaps as \textbf{design-system–implementation drifts}. Some drifts are unintentional deviations that should be corrected; others are legitimate evolution signals that the design system itself should absorb~\cite{overlayqa_design_system_drift}. Distinguishing the two is a software quality-management decision that requires design-system-level evidence.

Foundation-model-based UI generation~\cite{anthropic2026claudeDesign,bolt,stitch,vercelV0} amplifies this quality-management problem. Agent-driven changes can span dozens of files, multiple pages, and several component families in a single step, arriving faster than humans can review and before downstream agents consume them. Whether a change is human- or model-authored, reviewers face the same underlying question—"how does this change move the implemented design system, and should the team accept that move?"

Prior work proposed various solutions, but it does not directly recover these repository-level structured design system records. Visual UI quality-assurance techniques detect predefined display failures, accessibility violations, or deviations from explicit design guidance~\cite{su2021owleyes,su2022metamorphosis,chiou2024reflow,alshayban2022accessitext,krishnavajjala2024motorease,yang2021don,duan2024feedback,duan2024uicrit,yuan2025designrepair}. UI structure recovery, GUI grouping, and screenshot-to-code techniques identify elements or perceptual regions within individual screens~\cite{xie2020uied,chen2020object,xie2022psychologically,chen2023uilayers,li2022uldgnn,zhang2026guigroup,nguyen2015remaui,wan2025divide,wu2025mllm}. UI evolution and test-repair techniques establish element correspondences across versions~\cite{moran2018detecting,shao2021webevo,hammoudi2016waterfall,stocco2018visual,brisset2023sftm,salma2024guievo}. However, these works usually assume a predefined reference, operate at screen or element granularity, or target test maintenance rather than recovering functional groups, variants, and style conventions as repository-level records.

\improve{Recovering these relationships raises three challenges:}
First, the right analysis granularity sits between low-level implementation artifacts and whole-page screenshots: a newly added element may be a new component family, a variant of an existing family, or merely another occurrence of an established convention. Second, design-system-level consistency is not the same as visual identity. Two regions may look different while following the same project convention, and two regions may look similar while relying on different structures or style values. Third, recovered functional UI group and family records evolve across commits and pages, including newly added pages that have no previous same-page counterpart; lifecycle inference therefore requires matching functional groups and variants with style-value evidence across pages and versions rather than only aligning old and new elements.

We present \tool{}, a cross-artifact approach that recovers repository-level implemented design-system records and tracks its functional UI group and family lifecycle states evolve across commits. Given selected repository versions, \tool{} renders target pages and collects runtime evidence from screenshots, DOM hierarchies, UI-element boxes, and style records. It generates candidate regions from interactive containers, shared navigation targets, and DOM containers, and constrains a taxonomy-guided VLM to select, remove, or merge only these fixed candidates into functional UI groups. Finally, \tool{} aggregates recovered groups and style values across pages into repository-level UI group families and their variants, and repository-shared style values.
For every recovered family and functional UI group record, \tool{} also infers a lifecycle state across adjacent commits—Reused, Changed, Introduced, Removed, or Ambiguous (Table~\ref{tab:lifecycle_states})—giving human reviewers and downstream agents a design-system-level view of each change and a basis for deciding whether to absorb it into the design system or revert against it.

We evaluate \tool{} on our collected dataset of 10 open-source frontend repositories, covering 37 page timelines, 317 page-version instances, 280 adjacent-version transitions, 3,478 annotated functional UI groups, and 2,435 transition-level lifecycle states. \tool{} achieves an F1 of 0.85 for functional UI group recovery, outperforming direct VLM box generation and UI grouping baselines. For lifecycle inference, \tool{} achieves an overall lifecycle-state accuracy of 0.84, compared with 0.71 for the strongest baseline in our setting. Ablation results show that fixed candidate boxes, functional-category evidence, rendered style-change evidence, and repository-level style aggregation each contribute to the final performance. Our implementation and dataset are available at \url{https://github.com/Mingyue-eva/Designevo}.
This paper makes the following contributions:
\begin{itemize}
    \item We define a repository-level representation of implemented design systems that links functional UI groups, recurring families, variants, and shared style values to their page occurrences and cross-version correspondences, supporting traceable analysis of UI evolution.
    \item We introduce \tool{}, a cross-artifact approach that combines program-derived candidate regions with taxonomy-guided VLM judgments to recover functional UI groups, followed by repository-level aggregation and lifecycle-state inference to track their evolution across commits.
    \item We construct an annotated benchmark of 10 open-source frontend repositories and evaluate group recovery and lifecycle-state inference through baseline comparisons, ablations, and model-sensitivity analysis, demonstrating the contributions of constrained localization, functional categories, and style evidence.
\end{itemize}

\section{Related work}

\subsection{Automated UI Quality Assurance}

Automated UI quality assurance increasingly uses rendered interfaces to detect user-visible problems. OwlEye detects and localizes display failures~\cite{su2021owleyes}. dVermin identifies scaling failures by comparing interfaces under default and enlarged display scales~\cite{su2022metamorphosis}, while Chiou et al.\cite{chiou2024reflow} detect reflow accessibility issues in responsive web pages. AccessiText audits text-presentation requirements\cite{alshayban2022accessitext}, and MotorEase checks motor-accessible touch targets and spacing~\cite{krishnavajjala2024motorease}. These methods show the value of rendered evidence, but their judgments are tied to predefined defect classes, transformations, or accessibility rules.
Other work evaluates interfaces using broader design knowledge. Yang et al.\cite{yang2021don} and Seenomaly\cite{zhao2020seenomaly} operationalize Material Design guidance to detect visual design smells and problematic animation effects. Duan et al.\cite{duan2024feedback} generate feedback from written design heuristics, while UICrit\cite{duan2024uicrit} learns from expert critiques and quality ratings. UIClip~\cite{wu2024uiclip} learns general screenshot quality and relevance. DesignRepair~\cite{yuan2025designrepair} uses component and system-level Material Design knowledge to audit and revise frontend implementations, whereas GUIRepair~\cite{huang2025seeing} uses visual issue observations for fault comprehension and rerendered outputs for patch validation. These approaches support assessment or repair, but generally depend on explicit references such as guidelines, labeled critiques, bug reports, quality labels, or maintained design systems.

DesignEvo addresses a prerequisite for such auditing. In many repositories, project-specific design knowledge is implicit in recurring functional regions, reusable variants, shared style values, pages, files, and commits. Rather than judging whether a change is acceptable or defective, DesignEvo recovers these implemented repository-level design-system records for downstream comparison.

\subsection{Reverse Engineering for Design System Recovery}

We use design system recovery to mean recovering repository-level implemented design-system records: functional UI groups, UI group families, UI group variants, repository-shared style values, and lifecycle states. This target is related to reverse engineering, which reconstructs higher-level representations from system components and relationships~\cite{chikofsky1990reverse}. Prior UI and repository work has recovered interactive implementations, application storyboards, architectures, documentation, and traceability~\cite{hibschman2015unravel,chen2022automatically,zhang2023software,zhao2025software,keim2024recovering,luo2024repoagent}.
DesignEvo extends this line of work by lifting evidence from code, the DOM, rendered UIs, styles, and repository history into an implemented repository-level design system. \improve{Source-level reuse provides structural evidence, while the recovered families and variants describe how recurring functional patterns are actually rendered across pages; they need not coincide with the boundaries of a documented component library.}
Studies have cataloged refactorings for HTML, general Web applications, and React components~\cite{harold2012refactoring,nederlof2014software,ferreira2024refactoring}. Other techniques identify unused selectors, duplicated declarations, redundant rules, and overridden styles through static and dynamic analyses~\cite{mesbah2012automated,mazinanian2014discovering,hague2015detecting,punt2016pattern}. These analyses reveal shared style values but primarily support implementation maintenance rather than recovery.

DesignEvo also relates to UI structure recovery and frontend maintenance. Element-detection methods identify UI elements and merge fragmented visual layers~\cite{xie2020uied,chen2020object,chen2023uilayers,li2022uldgnn}, while perceptual-grouping~\cite{xie2022psychologically} and GUI grouping~\cite{zhang2026guigroup} infer regions from visual, spatial, semantic, or view-hierarchy cues~\cite{chen2024egfe,xiao2024ui,wen2024autodroid}. Screenshot-to-code systems use related decomposition to generate interfaces~\cite{nguyen2015remaui,wan2025divide,wu2025mllm}. Web and CSS analyses catalog HTML, Web, and React refactorings or detect unused selectors, duplicated declarations, redundant rules, and overridden styles~\cite{harold2012refactoring,nederlof2014software,ferreira2024refactoring,mesbah2012automated,mazinanian2014discovering,hague2015detecting,punt2016pattern}. These techniques expose structural or style evidence, but mainly target individual screens, code maintenance, duplication, or refactoring. DesignEvo combines these evidence types to recover UI group families and shared style values, interpreting repeated values through functional groups, variants, and usage contexts.

\subsection{UI Evolution Analysis}

Software artifacts continually evolve across commits and releases. Analyzing such evolution requires establishing correspondences between artifacts across versions despite additions, removals, renaming, layout changes, and refactoring. In the UI domain, GCAT summarizes GUI changes between mobile-app versions~\cite{moran2018detecting}, while GuiEvo supports mobile GUI evolution analysis~\cite{salma2024guievo}. WebEvo is the work most closely related to DesignEvo; it detects semantic structural changes in web pages and maps corresponding elements~\cite{shao2021webevo}.
UI test-repair techniques also establish element correspondences under interface evolution. WATER~\cite{choudhary2011water} and VISTA~\cite{stocco2018visual} match elements using attribute and visual information respectively. SFTM~\cite{brisset2023sftm} performs flexible tree matching. UITESTFIX~\cite{lin2023automated} extracts element information, while EDIT DIS~\cite{xu2025understanding} builds on this information and performs matching based only on XPath similarity. These approaches provide representative strategies for matching evolving UI elements, but their primary objective is maintaining test locators and test stability. More general software-evolution techniques track histories, refactorings, migrations, and cross-artifact consistency~\cite{liu2023automated,tsantalis2018accurate,hasan2024refactoring,ramos2023melt,keim2024recovering,tan2024detecting,pai2025codocbench,taha2026miningweb}, but typically focus on code entities, refactoring instances, migration rules, documentation links, or DOM changes.
DesignEvo builds on cross-version correspondence but lifts the analysis target from individual elements, page structures, and test locators to the lifecycle of implemented repository-level design-system records. \improve{Its hybrid analysis combines deterministic runtime evidence with constrained semantic judgment to relate page occurrences to functional families, variants, and shared conventions across the repository. This implemented representation supplies the evidence that a visual diff or a documented component catalog alone leaves unresolved.}

\section{Background}

\subsection{Design Systems in Frontend Implementations}
\label{sec:background_design_systems}

Modern web applications use design systems to coordinate interface decisions across products, teams, and releases. A design system typically comprises \emph{design tokens}, \emph{reusable components}, and \emph{interface patterns}, supported by implementation libraries, documentation, and usage guidelines. Design tokens encode shared visual decisions such as colors, typography, spacing, and shadows~\cite{design_tokens_format}. Reusable components encapsulate interface elements such as buttons, inputs, and cards, while interface patterns organize these elements into larger functional regions, such as forms, navigation panels, and dialogs~\cite{atomic_design_frost,pattern_lab}. Established systems, including Carbon, Ant Design, Material Design, and Fluent, illustrate how these elements are organized and exposed through documentation and component libraries~\cite{carbon_design_system,ant_design,material_design,fluent_design_system}.

In frontend implementations, these design decisions appear in different forms. Components may be instantiated through shared libraries, locally extended, or implemented independently using similar structures. Likewise, style values may be expressed through named tokens, CSS variables, utility classes, or hard-coded declarations. Consequently, the same functional pattern can have different source-code structures, while a shared component can produce different rendered variants depending on its properties, content, and context. Collectively, the recurring functional regions, component variants, and shared visual conventions expressed in an application constitute its \emph{implemented design system}.

As applications evolve, these implementations change through component modifications, local adaptations, style updates, and the introduction or removal of interface patterns. For example, a card may acquire additional controls, a form may adopt a different layout, or a shared color may be replaced by a locally defined value. Such changes can be intentional improvements or can introduce inconsistencies between intended conventions and their implementations, or among pages expected to follow the same conventions. We refer to these inconsistencies as \emph{design-system implementation drift}. Industry discussions distinguish token, component-variant, pattern, documentation, and behavioral drift~\cite{overlayqa_design_system_drift}. These categories reflect different aspects of design-system evolution, but identifying an inconsistency does not by itself establish whether the underlying change is defective or justified.

\subsection{Evidence for Recovering Implemented Design Systems}
\label{sec:existing_artifacts}

Understanding an implemented design system requires identifying recurring functional patterns, their rendered occurrences and style conventions, and how these relationships change across versions. Existing design artifacts provide complementary evidence, but differ in what they capture and how directly they represent the current implementation.

\textbf{Design files and documentation.}
Design files describe intended layouts, components, and visual properties, while design-system documentation specifies reusable elements, supported variants, and usage rules. For example, Carbon documents button variants and their intended purposes~\cite{carbon_button_usage}, and the Design Tokens Format Module standardizes the representation of named design values~\cite{design_tokens_format}. However, these artifacts do not necessarily identify where a pattern is instantiated in a particular application or how it has been locally adapted. Their use for implementation analysis therefore depends on their availability, synchronization with the analyzed version, and correspondence to actual pages and rendered states.

\textbf{Source-level components and styles.}
Frontend source code exposes component declarations, imports, properties, and style definitions. Such information can reveal explicit component reuse and shared style dependencies. However, functional equivalence does not always correspond to source-level identity: independently implemented components may serve the same purpose, while a shared component may render different variants depending on its properties and children~\cite{react_component_props}. Similarly, equivalent visual values may originate from named tokens, computed styles, or hard-coded declarations. Source-level relationships alone therefore do not fully characterize recurring patterns as users experience them.

\textbf{Rendered interfaces and visual changes.}
Screenshots and rendered page structures expose the appearance, composition, and spatial organization of implemented interfaces. Comparing versions can reveal additions, removals, and visual modifications. However, pixel-level differences alone do not establish whether changed regions serve the same function or belong to a recurring pattern across pages. UI evolution techniques can establish element correspondences across versions~\cite{shao2021webevo}, but such correspondences do not necessarily identify repository-level pattern families or their shared style conventions.

In summary, these artifacts provide different parts of the evidence needed to characterize an implemented design system. Recovering a repository-level view requires connecting functional regions to their rendered occurrences, identifying recurring patterns and shared styles across pages, and tracking their changes across versions. When existing documentation or tooling maintains such correspondences, it can directly support this analysis; otherwise, they must be recovered from the implementation. These recovered relationships characterize observable design-system evolution without independently determining whether a change conforms to an intended specification.


\begin{figure*}
    \centering
    \includegraphics[width=1\linewidth]{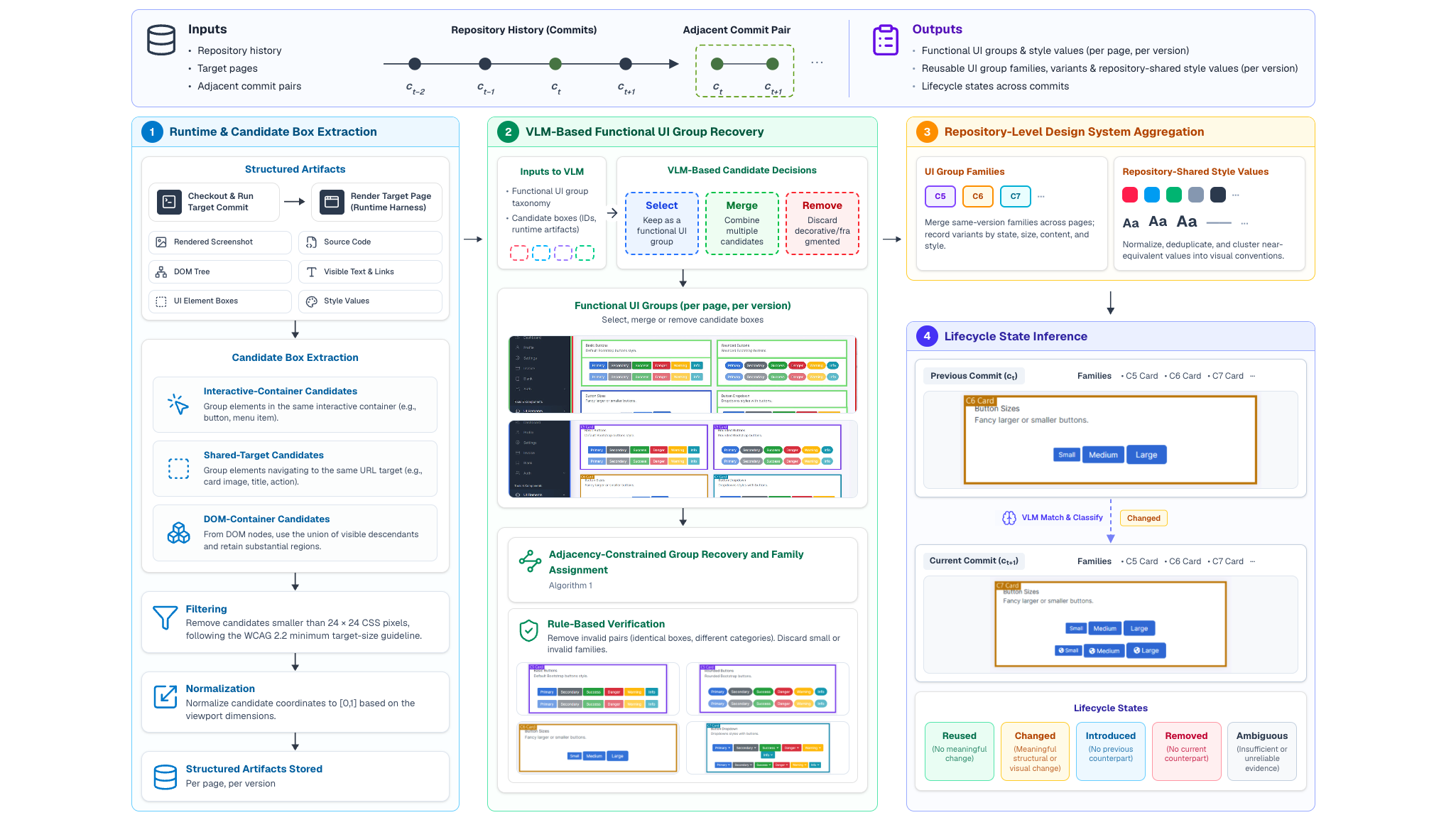}
    \caption{\textbf{Overview of DesignEvo.} DesignEvo takes repository histories, target pages, and adjacent commit pairs as input. For each page version, it renders the target commit, collects cross-artifact runtime evidence, and extracts fixed candidate boxes from interactive containers, shared URL targets, and DOM containers. A taxonomy-guided VLM then selects, merges, or removes these candidates to recover functional UI groups and same-version group families. DesignEvo aggregates recovered groups and style values into repository-level design-system records, and compares records across adjacent commits to infer \textsc{Reused}, \textsc{Changed}, \textsc{Introduced}, \textsc{Removed}, and \textsc{Ambiguous} lifecycle states.}
    \label{fig:method_overview}
\end{figure*}

\section{Methodology}
\label{sec:method}

DesignEvo recovers the project-specific \improve{implemented} design system implicit in a repository and tracks its evolution across commits. The key challenge is that such design knowledge is distributed across multiple artifacts: source code contains style declarations and reuses UI structures, DOM trees expose rendered UI element hierarchies, and rendered pages reveal the visual layout seen by users. 
No single artifact is sufficient, because code-level similarity may not imply visual or functional similarity, while visually similar UI groups may rely on different implementation structures. 
DesignEvo addresses this challenge through cross-artifact analysis over source code, DOM trees, and rendered pages. Its core design choice is to separate candidate-box generation from VLM-based semantic judgment: DesignEvo derives UI element boxes from rendered positions and sizes, and asks the VLM to identify functional UI groups from these fixed candidates rather than drawing boxes directly. The recovered groups, style values, and cross-commit correspondences are then aggregated into repository-level implemented design-system records.

\subsection{Overview}
\label{sec:method_overview}

As shown in Figure~\ref{fig:method_overview}, DesignEvo takes as input a repository history, a set of target pages, and adjacent commit pairs to compare. For each rendered page version, it recovers functional UI groups and their corresponding style values. For each repository version, it aggregates page-level records into repository-level implemented design-system records: UI group families, UI group variants, and repository-shared style values. For each adjacent commit pair, it outputs lifecycle states indicating whether these recovered records are introduced, reused, changed, removed, or ambiguous.
The pipeline has four stages. First, during runtime and candidate box extraction, the runtime harness renders pages; collects screenshots, source code, and style values; extracts the DOM tree, UI element boxes, visible text and links; and stores these structured artifacts. DesignEvo then extracts candidate boxes from three sources. Second, the VLM receives the Functional UI group taxonomy and the candidate boxes, and determines which candidates correspond to functional UI groups and how these UI groups should be interpreted. 
Third, DesignEvo aggregates page-level functional UI groups and style values into repository-level UI group families and repository-shared style values.
Fourth, DesignEvo compares recovered records across adjacent commits to produce cross-commit matching decisions and lifecycle states.

\subsection{Functional UI Group Taxonomy Construction}
\label{sec:method_taxonomy}

DesignEvo relies on a functional UI group taxonomy to guide VLM-based group recovery. The notion of a \textbf{functional UI group} is grounded in Atomic Design~\cite{atomic_design_frost}: it corresponds most closely to the organism level, where multiple UI elements and widgets are composed into a coherent interface region that serves a recognizable functional purpose. To make this unit operational for automated recovery, we construct a functional UI group taxonomy. The taxonomy is not intended to introduce new UI group types or enumerate all low-level UI elements and implementation components. Instead, it summarizes and merges common functional regions already present in design-system components, providing the VLM with an explicit taxonomy for page regions.

\textbf{Taxonomy construction.}
We constructed the taxonomy by reviewing public component catalogs and usage guidelines from nine widely used design systems listed by designsystems.surf~\cite{designsystems_surf}, including Apple, Atlassian, Ant Group, Amazon, IBM, Microsoft, Tailwind, Adobe, and Google. Component catalogs provide a practical starting point because they describe recurring interface units and their interaction purposes. 
Since design systems often use different names for functionally similar regions, we merged source names when they satisfy two criteria: \uline{they serve the same primary user-facing purpose, and they are typically perceived or edited as one coherent page region.} For example, the secondary label \textsc{Header} merges Header, App Bar, and Navigation Bar because all three denote a top-level region that groups branding, global navigation, and page-level controls. 

\textbf{Taxonomy structure.}
The resulting taxonomy contains nine primary categories: \textsc{Global Elements}, \textsc{Navigation}, \textsc{Action}, \textsc{Input}, \textsc{Data Display}, \textsc{Status}, \textsc{Overlay}, \textsc{Image Types}, and \textsc{Third-Party Embedded}. These primary categories contain secondary functional UI group labels, including \textsc{Header}, \textsc{Footer}, \textsc{Breadcrumb}, \textsc{Search}, \textsc{Sidebar}, \textsc{Action Bar}, \textsc{Input}, \textsc{Picker/Dropdown Selector}, \textsc{Form}, \textsc{Card}, \textsc{Stats Panel}, \textsc{Visualization}, \textsc{Timeline}, \textsc{Grid}, \textsc{Code Snippet}, \textsc{Data Table}, \textsc{List}, \textsc{Hero}, \textsc{Content Block}, \textsc{Message}, \textsc{Empty State}, \textsc{Loader}, \textsc{Progress}, \textsc{Dialog/Modal}, \textsc{Logo/Ad Block}, and \textsc{Embedded Widget}. We also report the mapping from the reviewed design-system component names to these taxonomy labels.
We provide the complete functional UI group taxonomy in the supplementary material submitted with this paper.

\textbf{Mapping and filtering criteria.}
A source name is left unmapped only when it primarily denotes a foundation concept, implementation utility, or low-level UI element rather than a functional UI group. Our criterion is whether the item usually forms a standalone functional region whose role can be compared across pages and commits. For example, Button, Icon, Label, Link, Badge, Slider, and Switch are included within functional UI groups and remain useful for analyzing UI element composition and style values, but they are not promoted to functional UI group labels. A single button or icon may change its style value, but its meaning is usually determined by the surrounding group,  such as \textsc{Search}, \textsc{Action Bar}, \textsc{Form}, \textsc{Card}, or \textsc{Dialog/Modal}. This mapping-and-filtering process covers common design-system components while keeping the label space compact and stable for UI group family recovery and lifecycle inference.

\subsection{Runtime and Candidate Box Extraction}
\label{sec:method_candidate_boxes}

For each target commit, DesignEvo checks out the repository, builds or runs the application under a configured runtime harness, and opens the target page. The runtime harness records  source code, a rendered page screenshot, DOM tree, visible text and links, UI element boxes and style values. This runtime view connects code-level changes to the rendered page: a code diff may show that a class name or style value changed, but rendering identifies the user-facing region affected by that change.
A candidate box is an automatically extracted region on a rendered page that may correspond to a functional UI group or part of one. DesignEvo extracts candidate boxes before invoking the VLM in the next step. This candidate set is intentionally broad and may contain overlapping or nested regions. because the next stage asks the VLM to select, remove, or merge candidates into functional UI groups.

DesignEvo extracts candidate boxes from three sources. \textbf{Interactive-container candidates} group visible UI elements that belong to the same interactive container, such as a button, link, menu item, or clickable card. \textbf{Shared-target candidates} group nearby visible UI elements that navigate to the same URL target, such as a card image, title, and action that all open the same detail page. \textbf{DOM-container candidates} recover functional UI groups that are not captured by shared interactive containers or shared URL targets. For each visible DOM node, DesignEvo computes an outer rectangle from the union of its visible descendant UI element boxes and retains the node as a candidate when the rectangle encloses a non-trivial visual region with sufficient area, visible content, or interaction targets. This rule covers regions such as forms, search areas, table regions, chart panels, headers, sidebars, and content cards, whose elements may jointly serve one function without sharing a single interactive container or navigation target.
Furthermore, DesignEvo filters out candidates smaller than \(24 \times 24\) CSS pixels, following the WCAG 2.2 minimum target-size guideline for pointer inputs~\cite{wcag22}, and normalizes their coordinates to [0,1] based on the viewport dimensions.

\subsection{VLM-Based Functional UI Group Recovery}
\label{sec:method_functional_group_recovery}

DesignEvo combines fixed candidate-box extraction with \textbf{VLM-based candidate decisions}. This design avoids asking the VLM to perform open-ended visual localization, which can produce unstable or non-reproducible coordinates. Instead, DesignEvo first extracts fixed candidate boxes from the rendered page, and the VLM only decides whether to \textsc{Select} each candidate as a functional UI group, \textsc{Remove} it, or \textsc{Merge} it with nearby candidates. This separates coordinate generation from functional interpretation and makes each recovered group traceable to concrete runtime records.

Given the candidate boxes and runtime harness records, the VLM recovers functional UI groups using the taxonomy defined in Section~\ref{sec:method_taxonomy}. For each decision, the VLM returns selected candidate IDs, a functional category, and a short reason. It can only refer to existing candidate IDs. If several candidate boxes jointly form one visible and functionally coherent region, the VLM outputs one functional UI group containing all corresponding candidate IDs. If a candidate box is decorative, duplicated, or only an incomplete fragment of a larger region, the VLM marks it as \textsc{Removed}. The program then materializes the final bounding box of each functional UI group as the union of its selected candidate boxes. The VLM never creates new coordinates and never edits box values.

\textbf{Adjacency-Constrained Group Recovery and Family Assignment.}
Algorithm~\ref{alg:functional_group_recovery} first recovers functional UI groups from candidate boxes using an adjacency graph that connects boxes that overlap, touch, or have no visual gap. For each connected component, the VLM decides whether to keep a candidate, merge multiple candidates, or remove decorative, duplicated, or incomplete fragments. DesignEvo materializes kept and merged candidates as functional UI groups using the union of their boxes.
It then assigns groups within the same repository version to UI group families. Rather than grouping all groups globally by functional category, DesignEvo builds a neighbor graph over spatially adjacent groups with compatible categories and expands families using BFS. For each neighboring pair, the VLM determines whether they are UI group variants of the same family based on their functional category, layout, UI element composition, style values, text, and surrounding context. This local strategy preserves likely merge relationships while avoiding comparisons across unrelated page regions.

\textbf{Rule-based verification.}
DesignEvo applies rule-based verification to reduce VLM hallucinations. It removes invalid pair decisions where two groups have identical bounding boxes but different functional categories, which usually indicate duplicate interpretations of the same region rather than variants of one UI group family. If removing such pairs disconnects a tentative family or leaves fewer than two groups, DesignEvo discards that family assignment.

\begin{algorithm}[t]
\caption{Adjacency-constrained functional UI group recovery and family assignment}
\label{alg:functional_group_recovery}
\scriptsize
\begin{algorithmic}[1]
\Require Rendered page screenshot $I$, candidate boxes $B$, runtime records $R$, functional UI group taxonomy $T$
\Ensure Functional UI groups $G$ and same-version UI group family assignments $F$

\State $G \gets \emptyset$
\State Construct an adjacency graph $A$ over candidate boxes in $B$
\Comment{Edges connect boxes that overlap, touch, or have no visual gap.}

\For{each connected component $C$ in $A$}
    \State $D_C \gets$ VLM decisions over candidates in $C$
    \Comment{Each decision is \textsc{Select}, \textsc{Merge}, or \textsc{Remove}.}
    \For{each decision $d \in D_C$}
        \If{$d$ is \textsc{Select} or \textsc{Merge}}
            \State $S_d \gets$ candidate IDs selected by $d$
            \State Materialize a functional UI group $g_d$ from $S_d$
            \State Set $g_d$'s bounding box to the union of boxes in $S_d$
            \State Assign $g_d$ the functional category returned by the VLM
            \State Add $g_d$ to $G$
        \EndIf
    \EndFor
\EndFor

\State Construct a neighbor graph $N$ over $G$
\Comment{Edges connect spatially adjacent groups with compatible functional categories.}
\State $F \gets \emptyset$
\State $Assigned \gets \emptyset$

\For{each group $g \in G$}
    \If{$g \notin Assigned$}
        \State Initialize tentative family $f \gets \{g\}$ and accepted edges $E_f \gets \emptyset$
        \State Initialize a BFS queue $Q \gets [g]$
        \While{the queue is not empty}
            \State Pop a group $u$ from $Q$
            \For{each neighbor $v$ of $u$ in $N$ where $v \notin f$ and $v \notin Assigned$}
                \State Ask the VLM whether $u$ and $v$ are UI group variants of the same UI group family
                \If{the VLM accepts the pair}
                    \State Add $v$ to $f$ and add edge $(u,v)$ to $E_f$
                    \State Push $v$ into $Q$
                \EndIf
            \EndFor
        \EndWhile
    \EndIf
\EndFor

\State \Return $G, F$
\end{algorithmic}
\end{algorithm}

\subsection{Repository-Level Design System Aggregation}
\label{sec:method_repository_aggregation}

DesignEvo aggregates page-level recovery results into repository-level implemented design-system records for each repository version. These records consist of two complementary parts: UI group families and repository-shared style values. Together, they approximate the implemented design system evidenced in the repository because they capture both what functional UI groups are reused across pages and which style values are used to express their visual appearance. 

\textbf{UI group family aggregation.}
DesignEvo uses the VLM to merge same-version family assignments across target pages into UI group families. Groups are merged when they have compatible functional categories, the same primary user-facing purpose, and a recognizable shared structure or visual convention. Differences in state, size, content, or visual presentation are recorded as UI group variants rather than separate families.

\textbf{Repository-shared style value aggregation.}
DesignEvo aggregates group-level style values into repository-level style records through three steps. First, it applies \emph{frequency weighting} to identify values that recur across recovered functional UI groups and reduce the influence of incidental page-specific styles. Second, it performs \emph{near-equivalent grouping} for values under the same style property, such as perceptually similar colors, adjacent spacing values, or equivalent radius values expressed through different literals or variables. Third, it applies \emph{taxonomy-based naming} to map recurring value groups to shared visual conventions, such as color palettes, spacing scales, typography scales, and radius scales.
The aggregation preserves both normalized repository-level records and the underlying implementation records, including CSS class names, declared values when resolvable, computed values at the rendered viewport, and associated functional UI groups. 

\subsection{Lifecycle State Inference}
\label{sec:method_lifecycle}

After recovering repository-level design system records for each version, DesignEvo infers lifecycle states to make these records usable for UI change auditing. A repository-level record that is introduced, reused, changed, removed, or ambiguous across commits provides evidence about whether a UI change follows or departs from the project's evolving conventions. DesignEvo therefore compares repository-level records across adjacent commits and assigns one of the lifecycle states in Table~\ref{tab:lifecycle_states}. These records include UI group families and repository-shared style values.
The VLM then classifies each candidate match into \textsc{Reused}, \textsc{Changed}, or \textsc{Ambiguous}. DesignEvo marks unmatched current records as \textsc{Introduced} and unmatched previous records as \textsc{Removed}.
We use \textsc{Changed} for cases that preserve the identity of a repository-level record but alter its rendered structure, UI element composition, style values, or shared visual convention. If multiple previous UI group families correspond to one current family, or one previous family corresponds to multiple current families, DesignEvo still uses \textsc{Changed} when the role and primary user-facing purpose are preserved, and records the merge or split in the reason. This keeps the lifecycle state space compact and avoids introducing rare states that increase classification complexity.

\begin{table}[t]
\centering
\caption{Lifecycle states used by DesignEvo.}
\label{tab:lifecycle_states}
\scriptsize
\setlength{\tabcolsep}{4pt}
\begin{tabular}{p{0.27\columnwidth}p{0.63\columnwidth}}
\toprule
\textbf{State} & \textbf{Definition} \\
\midrule
\textsc{Reused} & Same repository-level record without meaningful visual, structural, or style value change. \\
\textsc{Changed} & Same repository-level record with meaningful layout, size, style value, content arrangement, UI element composition, or convention-level change. \\
\textsc{Introduced} & Appears in the current version without a previous counterpart. \\
\textsc{Removed} & Existed in the previous version but has no current counterpart. \\
\textsc{Ambiguous} & Noisy, unreliable, or insufficiently supported correspondence. \\
\bottomrule
\end{tabular}
\end{table}

\section{Experiments}
\label{sec:experiments}

\subsection{Subjects}

\begin{table*}[t]
\centering
\caption{Repository-level composition of the DesignEvo dataset.}
\label{tab:repository_composition}
\scriptsize
\setlength{\tabcolsep}{3pt}
\begin{tabular*}{\linewidth}{@{\extracolsep{\fill}}>{\raggedright\arraybackslash}p{0.34\linewidth}>{\raggedright\arraybackslash}p{0.14\linewidth}rrrrr@{}}
\toprule
\textbf{Repository} &
\textbf{Primary stack} &
\textbf{Timelines} &
\textbf{Instances} &
\textbf{Transitions} &
\textbf{\makecell[r]{Functional\\UI Groups}} &
\textbf{\makecell[r]{Lifecycle\\State Labels}} \\
\midrule
\texttt{BootstrapDash/\allowbreak PurpleAdmin-\allowbreak Free-\allowbreak Admin-\allowbreak Template}
& Bootstrap & 3 & 21 & 18 & 221 & 147 \\
\texttt{ColorlibHQ/\allowbreak AdminLTE} (v3 branch)
& Bootstrap/\allowbreak jQuery & 5 & 38 & 33 & 500 & 364 \\
\texttt{StartBootstrap/\allowbreak startbootstrap-\allowbreak sb-\allowbreak admin-\allowbreak 2}
& Bootstrap & 4 & 39 & 35 & 292 & 249 \\
\texttt{YafetGetu/\allowbreak Admin-\allowbreak Dashbored}
& Django/\allowbreak HTML & 1 & 11 & 10 & 75 & 76 \\
\texttt{adminkit/\allowbreak adminkit}
& Bootstrap & 5 & 40 & 35 & 406 & 261 \\
\texttt{creativetimofficial/\allowbreak soft-\allowbreak ui-\allowbreak dashboard}
& Bootstrap & 2 & 16 & 14 & 144 & 104 \\
\texttt{design-\allowbreak sparx/\allowbreak antd-\allowbreak multipurpose-\allowbreak dashboard}
& React/\allowbreak Ant Design & 5 & 20 & 15 & 361 & 209 \\
\texttt{puikinsh/\allowbreak CoolAdmin}
& Bootstrap & 4 & 59 & 55 & 554 & 458 \\
\texttt{themeselection/\allowbreak sneat-\allowbreak bootstrap-\allowbreak html-\allowbreak admin-\allowbreak template-\allowbreak free}
& Bootstrap & 4 & 31 & 27 & 675 & 399 \\
\texttt{zuramai/\allowbreak mazer}
& Bootstrap & 4 & 42 & 38 & 250 & 168 \\
\midrule
\textbf{Total}
& --
& \textbf{37}
& \textbf{317}
& \textbf{280}
& \textbf{3,478}
& \textbf{2,435} \\
\bottomrule
\end{tabular*}
\end{table*}

\textbf{Repository and Version Collection.}
We constructed the \textbf{DesignEvo dataset} from 10 open-source frontend repositories spanning different application domains, project sizes, and representative frontend stacks, including Bootstrap, React, Ant Design, Tailwind CSS, and Django. The selected repositories contain component-rich, multi-page interfaces and recurring UI structures, making them suitable for studying repository-level functional UI group reuse and evolution.

For each repository, we retrieved its commit history and retained commits containing substantive modifications to frontend UI files. We excluded changes limited to documentation, configuration, dependencies, comments, formatting, or textual content, and filtered near-duplicate commits with negligible UI differences.
We then checked out each retained version and attempted to build and render its accessible application routes. Versions that could not be reproducibly installed, launched, or rendered were excluded. Each successfully rendered route at a commit forms a \emph{page-version instance}; chronologically ordered instances of the same route form a \emph{page timeline}, and adjacent valid instances define the \emph{version transitions} used in our analysis. We retained only routes with at least two valid rendered versions and observable UI changes.
For each page-version instance, we collected its screenshot, DOM structure, computed styles, element geometry, route, and commit metadata. For adjacent versions, we additionally generated visual and structural comparisons to support annotation and baseline construction. These automatically generated artifacts served only as evidence; all final annotations were manually reviewed.

Annotators reviewed the before-and-after renderings together with the corresponding DOM and change evidence. They identified functional UI groups, assigned functional categories, and labeled each group’s lifecycle state as \textsc{Reused}, \textsc{Changed}, \textsc{Introduced}, \textsc{Removed}, or, when the available evidence was insufficient, \textsc{Ambiguous}. Groups first observed at the beginning of a page timeline were labeled \textsc{Introduced}.
Two annotators independently reviewed the generated comparison instances, and 20\% of the instances were re-annotated for quality control. Their agreement exceeded 97\%. Disagreements were resolved through discussion and third-person adjudication; instances that remained insufficiently defined were excluded from the final dataset.

\textbf{Dataset Statistics.}
The final dataset contains 10 repositories and 37 page timelines, comprising 317 page-version instances and 280 version transitions. Across these instances and transitions, we annotated 3,478 functional UI groups and 2,435 transition-level lifecycle-state labels. Table~\ref{tab:repository_composition} summarizes the contribution of each repository.
The number of functional UI groups contributed by individual repositories ranges from 75 to 675, and the number of lifecycle-state labels ranges from 76 to 458, reflecting differences in the number, length, and complexity of collected page timelines. Specifically, \textsc{Reused} labels account for 71.01\% of all lifecycle-state annotations. The remaining 28.99\% capture observable UI evolution, including \textsc{Changed} (15.56\%), \textsc{Introduced} (7.06\%), \textsc{Removed} (6.24\%), and \textsc{Ambiguous} (0.12\%) labels, following the lifecycle definitions in Table~\ref{tab:lifecycle_states}. The dataset therefore contains both persistent UI structures and cross-version changes for evaluating functional UI group matching and lifecycle-state inference. 
Detailed distributions of primary and secondary functional UI group types are provided in the supplementary material.

\subsection{Research Questions}
\noindent We evaluate DesignEvo through three research questions. Each RQ targets a specific design choice in the methodology before reporting the corresponding setup and results.

\begin{itemize}
    \item \textbf{RQ1: Functional UI group recovery.}
    How accurately can DesignEvo recover functional UI groups from one repository commit, and compared with baseline methods?
    \item \textbf{RQ2: Lifecycle-state inference.}
    How accurately can DesignEvo infer lifecycle-state labels for recovered functional UI group and family records across adjacent commits and compared with baseline methods?
    \item \textbf{RQ3: Ablation of core evidence and constraints.}
    How much do DesignEvo's core evidence sources and pipeline constraints contribute to functional UI group recovery, repository-level design system record construction, and lifecycle-state inference?
\end{itemize}

\subsection{Shared Experimental Protocol}
\label{sec:shared_protocol}

\textbf{VLM configuration.}
The main experiments use the GPT-5.5 VLM recorded in our experiment logs because it supports image understanding, structured JSON output, and candidate-box semantic judgment. We set temperature to 0, use the gateway-default top-$p$, and set task-specific token limits for JSON outputs. Prompts and raw responses are retained for auditing. Since DesignEvo relies on VLM judgments for functional-group recovery, family assignment, and lifecycle-state inference, we also run a model-sensitivity study with Qwen2.5-VL for RQ1 and RQ2 using the same inputs, prompts, candidate sets, and scoring scripts.

\textbf{Evaluation metrics.}
For RQ1, we greedily match predicted and reference boxes by descending IoU. A prediction is counted as a true positive only when it matches a reference box with IoU $\geq 0.9$ and the correct secondary functional label. Unmatched predictions and label-mismatched matches are false positives; unmatched references are false negatives. We report precision, recall, F1, and mean bbox IoU, where mean bbox IoU is computed only over true-positive matches.

For lifecycle-state inference in RQ2 and RQ3, we report per-state accuracy when each method produces one label per annotated state, and precision/recall/F1 when methods may produce extra or missing records. All aggregate precision, recall, and F1 scores are micro-averaged from exact counts over the full evaluation universe. Comparable methods use the same annotated page versions, transitions, candidate correspondences, and scoring scripts.

\subsection{RQ1: Functional UI Group Recovery}
\label{sec:rq1}

RQ1 evaluates DesignEvo’s separation of candidate extraction from VLM-based functional judgment. Because downstream aggregation and lifecycle inference depend on accurate functional UI group recovery, RQ1 tests whether constrained VLM judgment over fixed candidates outperforms unconstrained box generation and prior UI grouping methods.

\subsubsection{\textbf{Experimental Setup}}
\label{sec:rq1_protocol}

We evaluate RQ1 using 3,478 manually annotated functional UI groups and the matching protocol described in Section~\ref{sec:shared_protocol}. We compare DesignEvo (GPT-5.5) against three baselines and one model-sensitivity variant. \textbf{Unconstrained VLM Localization (GPT-5.5)} removes candidate box extraction and asks the same VLM to directly localize functional UI groups. \textbf{GUI Perceptual Grouping}\cite{xie2022psychologically} represents a psychology-inspired, rule-based visual grouping approach, whereas \textbf{GUIGroup}\cite{zhang2026guigroup} represents an MLLM-based functional grouping baseline. \textbf{DesignEvo (Qwen2.5-VL)} retains DesignEvo’s pipeline, prompts, candidate boxes, and scoring scripts while replacing only the model.

\subsubsection{\textbf{Results}}

\begin{figure*}[t]
\centering
\captionsetup[subfigure]{font=scriptsize,justification=centering}
\setlength{\abovecaptionskip}{2pt}
\begin{subfigure}[t]{0.49\textwidth}
    \centering
    \includegraphics[width=\linewidth]{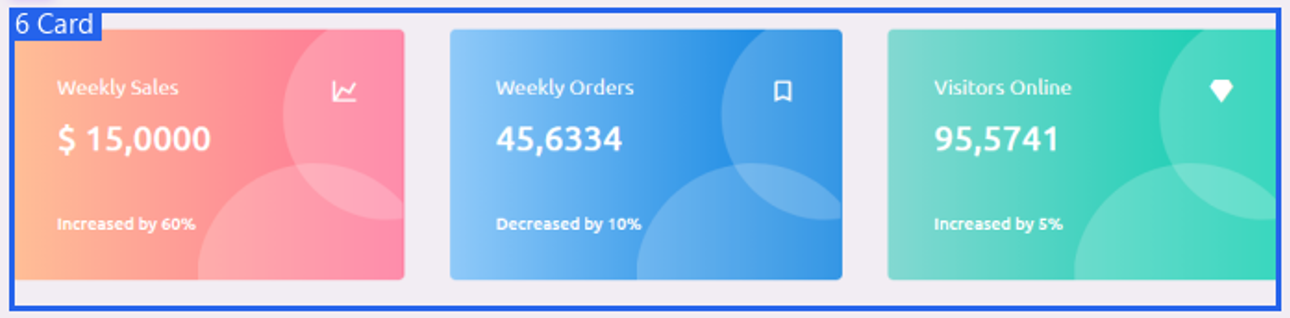}
    \caption{GUI grouping merges repeated cards.}
\end{subfigure}
\hfill
\begin{subfigure}[t]{0.49\textwidth}
    \centering
    \includegraphics[width=\linewidth]{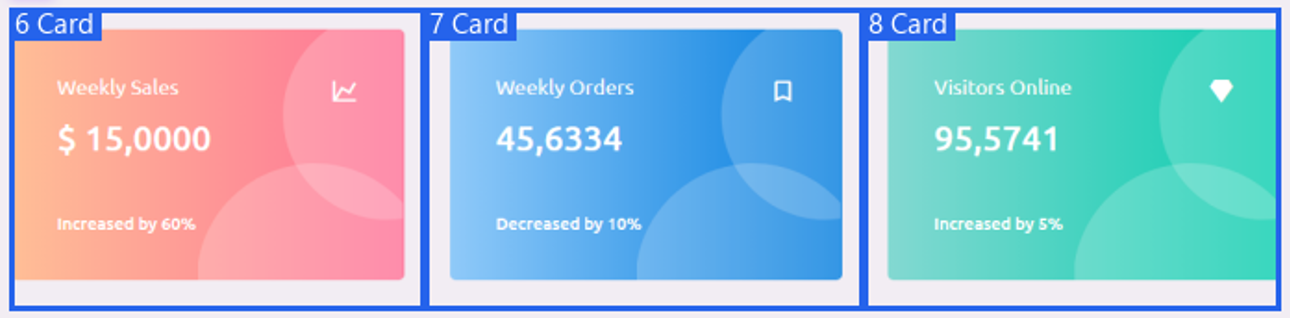}
    \caption{DesignEvo keeps separate card-variant groups.}
\end{subfigure}
\begin{subfigure}[t]{0.49\textwidth}
    \centering
    \includegraphics[width=\linewidth]{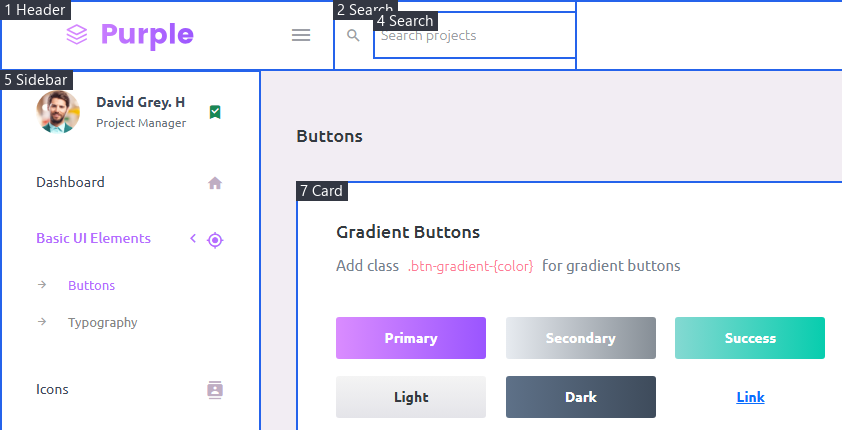}
    \caption{Qwen2.5-VL leaves a search group unmerged.}
\end{subfigure}
\hfill
\begin{subfigure}[t]{0.49\textwidth}
    \centering
    \includegraphics[width=\linewidth]{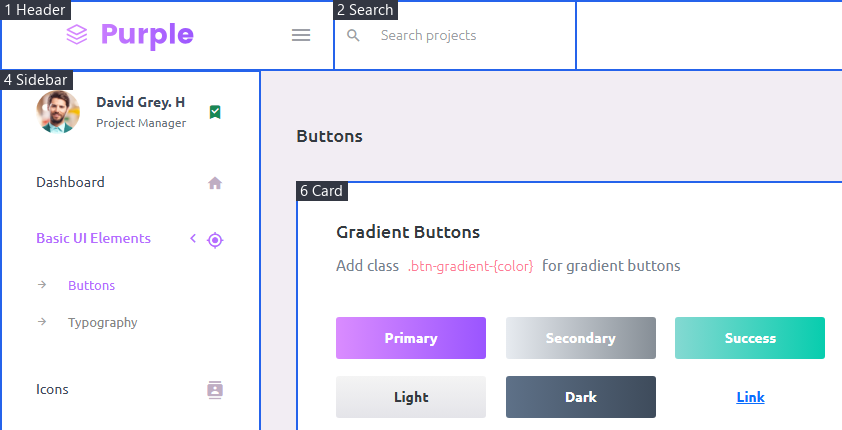}
    \caption{GPT-5.5 merges the related candidates.}
\end{subfigure}
\caption{Compact qualitative examples for RQ1. The first row compares functional UI group recovery, the second row shows model sensitivity in candidate merging.}
\label{fig:rq_examples}
\end{figure*}

Table~\ref{tab:rq1_group_recovery} shows that DesignEvo (GPT-5.5) achieves the best functional UI group recovery accuracy, with an F1 of 0.8510 and mIoU of 0.9791. Unconstrained VLM Localization (GPT-5.5), which asks the same VLM to directly draw group boxes, drops to an F1 of 0.4223 and recall of 0.3470, showing that open-ended localization is unreliable for this task. GUI perceptual grouping and GUIGroup perform better than unconstrained localization but remain below DesignEvo, with F1 scores of 0.6683 and 0.7664. DesignEvo (Qwen2.5-VL) reaches an F1 of 0.7362: 
The fixed candidate boxes maintain consistent localization, but weaker merge and removal decisions reduce precision to 0.6866.
The 95\% CIs for DesignEvo are tight around the point estimates: precision 0.8538 [0.8408, 0.8647] and recall 0.8482 [0.8349, 0.8589]. Against the strongest baseline, GUIGroup, the paired improvement remains positive for both precision ([0.0351, 0.0579]) and recall ([0.0983, 0.1462]).
The qualitative cases in Figure~\ref{fig:rq_examples} explain the main different mechanisms. GUI perceptual grouping and GUIGroup may merge several repeated statistic cards into one large connected region, whereas DesignEvo keeps the cards as separate functional UI groups and assigns them to the same card variant. The model-sensitivity example further shows that the fixed candidate set does not remove all semantic judgment difficulty: Qwen2.5-VL leaves a search-related region unmerged, while GPT-5.5 correctly merges the related candidates into one functional group.

\begin{table}[t]
\centering
\caption{Accuracy of functional UI group recovery.}
\label{tab:rq1_group_recovery}
\scriptsize
\setlength{\tabcolsep}{4pt}
\begin{tabular}{lrrrr}
\toprule
\textbf{Method} & \textbf{Prec.} & \textbf{Rec.} & \textbf{F1} & \textbf{mIoU} \\
\midrule
GUI perceptual grouping~\cite{xie2022psychologically} & 0.7234 & 0.6210 & 0.6683 & 0.9365 \\
GUIGroup~\cite{zhang2026guigroup} & 0.8099 & 0.7274 & 0.7664 & 0.9451 \\
\makecell[l]{Unconstrained VLM Localization (GPT-5.5)} & 0.5393 & 0.3470 & 0.4223 & 0.8656 \\
DesignEvo (Qwen2.5-VL) & 0.6866 & 0.7936 & 0.7362 & 0.9621 \\
DesignEvo (GPT-5.5) & 0.8538 & 0.8482 & 0.8510 & 0.9791 \\
\bottomrule
\end{tabular}
\end{table}

\subsection{RQ2: Lifecycle-State Inference}
\label{sec:rq2}

RQ2 evaluates DesignEvo's lifecycle inference across adjacent commits. Rather than matching isolated DOM elements, DesignEvo compares recovered functional UI group families. RQ2 tests whether this repository-level representation improves lifecycle-state labeling over element-level matching baselines.

\subsubsection{\textbf{Experimental Setup}}
\label{sec:rq2_protocol}

We compare \textbf{DesignEvo (GPT-5.5)} with \textbf{WebEvo}, the closest related web-page evolution technique, which detects semantic structural changes and maps corresponding elements~\cite{shao2021webevo}. We also include three correspondence methods from UI test repair: \textbf{WATER} uses attribute-based matching~\cite{choudhary2011water}, \textbf{VISTA} uses visual matching~\cite{stocco2018visual}, and \textbf{SFTM} performs flexible tree matching~\cite{brisset2023sftm}. We further include two baselines derived from the same prior work~\cite{lin2023automated}: \textbf{EDIT DIS}, which matches elements using XPath edit distance, and \textbf{EDIT DIS+GPT-5.5}, which additionally applies GPT-5.5-based filtering and lifecycle-state classification. Finally, we report \textbf{DesignEvo (Qwen2.5-VL)} as a model-sensitivity variant that uses the same settings.

\subsubsection{\textbf{Results}}

Table~\ref{tab:rq2_lifecycle_accuracy} shows that DesignEvo (GPT-5.5) achieves the highest overall lifecycle-state accuracy, 0.84, compared with 0.71 for WebEvo and 0.62 for Edit Distance+GPT-5.5. The largest gains occur for states that require design-system-level interpretation rather than local element matching. For \textsc{Changed}, DesignEvo reaches 0.83 accuracy, compared with 0.72 for WebEvo. For \textsc{Introduced} and \textsc{Removed}, DesignEvo reaches 0.89 and 0.88, while WebEvo reaches 0.61 and 0.36. DesignEvo (Qwen2.5-VL) reaches 0.7415 overall accuracy as a model-sensitivity reference. These results indicate that repository-level UI group families and shared style values provide useful evidence for distinguishing true lifecycle changes from visual or DOM.

\begin{table}[t]
\centering
\caption{Accuracy of cross-commit lifecycle-state inference. DEvo-Qwen uses Qwen2.5-VL; EDIT DIS+GPT and DEvo GPT use GPT-5.5.}
\label{tab:rq2_lifecycle_accuracy}
\scriptsize
\setlength{\tabcolsep}{1.5pt}
\begin{tabular}{lrrrrrrrr}
\toprule
\textbf{State}
& \textbf{SFTM}
& \textbf{VISTA}
& \textbf{WATER}
& \textbf{WebEvo}
& \textbf{\makecell{EDIT\\DIS}}
& \textbf{\makecell{EDIT \\DIS+GPT}}
& \textbf{\makecell{DEvo\\Qwen}}
& \textbf{\makecell{DEvo\\GPT}} \\
\midrule
\textsc{Reused}     & 0.68 & 0.73 & 0.75 & 0.76 & 0.66 & 0.72 & 0.72 & 0.82 \\
\textsc{Changed}    & 0.62 & 0.71 & 0.65 & 0.72 & 0.54 & 0.53 & 0.72 & 0.83 \\
\textsc{Introduced} & 0.56 & 0.60 & 0.58 & 0.61 & 0.44 & 0.52 & 0.86 & 0.89 \\
\textsc{Removed}    & 0.21 & 0.29 & 0.24 & 0.36 & 0.17 & 0.26 & 0.77 & 0.88 \\
\textsc{Ambiguous}  & 0.00 & 0.33 & 0.00 & 0.50 & 0.00 & 0.00 & 0.00 & 0.50 \\
\midrule
\textsc{Overall}    & 0.62 & 0.68 & 0.68 & 0.71 & 0.53 & 0.62 & 0.74 & 0.84 \\
\bottomrule
\end{tabular}
\end{table}


\subsection{RQ3: Ablation of Core Evidence and Constraints}
\label{sec:rq3}

RQ3 evaluates the contribution of DesignEvo's core evidence sources and pipeline constraints. Because functional UI group recovery, repository-level record construction, and lifecycle-state inference depend on different forms of geometric, semantic, visual, and style evidence, end-to-end accuracy alone cannot show which parts of the pipeline are necessary. RQ3 therefore tests how accuracy changes when each major evidence source or constraint is removed.

\subsubsection{\textbf{Experimental Setup}}
\label{sec:rq3_protocol}

We evaluate four groups of ablations. \textbf{RQ3.1} tests candidate box extraction for functional UI group recovery by removing Interactive-container candidates, Shared-target candidates or DOM-container candidates.
\textbf{RQ3.2} tests same-version UI group family assignment by removing functional taxonomy evidence, bounding-box size and position evidence, or the rule-based verification guard. \textbf{RQ3.3} tests repository-shared style value aggregation by removing frequency weighting, near-equivalent value grouping, or taxonomy-based naming. 
\textbf{RQ3.4} tests lifecycle inference by removing functional-category evidence or repository-shared style-change evidence from cross-commit matching.


\subsubsection{\textbf{Results}}

\textbf{RQ3.1: Candidate box extraction.}
Table~\ref{tab:rq3_candidate_box_ablation} shows that fixed candidate boxes provide the largest observed contribution to functional UI group recovery. Allowing the VLM to produce boxes directly reduces F1 from 0.8510 to 0.4223 and recall from 0.8482 to 0.3470. This confirms the core RQ1 mechanism: unconstrained visual localization entangles the task of finding coordinates with the task of judging functional meaning. Removing shared-target grouping or shared-interactive-container grouping reduces F1 to 0.5802 and 0.5827, respectively. These comparable drops show that both grouping rules prevent functional UI groups from being recovered as fragmented regions. 

\begin{table}[t]
\centering
\caption{Ablation results for candidate box extraction.}
\label{tab:rq3_candidate_box_ablation}
\scriptsize
\setlength{\tabcolsep}{3.5pt}
\begin{tabular}{lrrrr}
\toprule
\textbf{Variant} & \textbf{Prec.} & \textbf{Rec.} & \textbf{F1} & \textbf{mIoU} \\
\midrule
DesignEvo & 0.8538 & 0.8482 & 0.8510 & 0.9791 \\
\makecell[l]{Unconstrained VLM\\Localization (GPT-5.5)} & 0.5393 & 0.3470 & 0.4223 & 0.8656 \\
Without shared-target grouping & 0.5918 & 0.5690 & 0.5802 & 0.9317 \\
Without shared-interactive-container grouping & 0.5894 & 0.5762 & 0.5827 & 0.9321 \\
\bottomrule
\end{tabular}
\end{table}

\textbf{RQ3.2: Same-version UI group family assignment.}
Table~\ref{tab:rq3_family_assignment_ablation} shows that bounding-box size and position evidence contribute most to suppressing false-positive family assignments, reducing precision from 0.9452 to 0.8555 when removed. This happens because visually similar cards, filters, and summary panels may share colors or typography but play different roles in different page regions; spatial evidence helps prevent these local similarities from becoming false family links. Removing taxonomy evidence reduces F1 from 0.8536 to 0.8406, showing that functional categories help distinguish visually similar groups with different roles. Removing rule-based verification produces a smaller F1 drop to 0.8495. Recall remains unchanged across these variants, while the number of predicted families increases, indicating that these constraints mainly improve precision by rejecting over-broad family assignments.

\begin{table}[t]
\centering
\caption{Ablation results for same-version UI group family assignment.}
\label{tab:rq3_family_assignment_ablation}
\scriptsize
\setlength{\tabcolsep}{3.5pt}
\begin{tabular}{lrrr}
\toprule
\textbf{Variant} & \textbf{Prec.} & \textbf{Rec.} & \textbf{F1} \\
\midrule
DesignEvo & 0.9452 & 0.7782 & 0.8536 \\
Without taxonomy & 0.9138 & 0.7782 & 0.8406 \\
Without bbox/size/position & 0.8555 & 0.7782 & 0.8150 \\
Without rule-based verification & 0.9351 & 0.7782 & 0.8495 \\
\bottomrule
\end{tabular}
\end{table}

\textbf{RQ3.3: Repository-shared style value aggregation.}
Table~\ref{tab:rq3_style_value_ablation} shows that the full three-step aggregation achieves an F1 of 0.9873. Removing frequency weighting causes the largest degradation, reducing precision to 0.4632 and F1 to 0.6332, because many incidental or page-specific values are no longer pruned. This is the strongest evidence that repository-level style recovery cannot simply union all observed computed values: one-off widths, local margins, chart colors, and content-specific dimensions would be misreported as shared design-system values. Removing near-equivalent value grouping reduces F1 to 0.8451, showing that equivalent conventions are often expressed through slightly different literals or variables. Removing taxonomy-based naming has a smaller effect, reducing recall to 0.9704 and F1 to 0.9796, which suggests that naming mainly affects coverage and interpretability.

\begin{table}[t]
\centering
\caption{Ablation results for repository-shared style value aggregation.}
\label{tab:rq3_style_value_ablation}
\scriptsize
\setlength{\tabcolsep}{4pt}
\begin{tabular}{lrrr}
\toprule
\textbf{Variant} & \textbf{Prec.} & \textbf{Rec.} & \textbf{F1} \\
\midrule
DesignEvo & 0.9750 & 1.0000 & 0.9873 \\
Without frequency weighting & 0.4632 & 1.0000 & 0.6332 \\
Without near-equivalent grouping & 0.7835 & 0.9172 & 0.8451 \\
Without taxonomy-based naming & 0.9889 & 0.9704 & 0.9796 \\
\bottomrule
\end{tabular}
\end{table}

\textbf{RQ3.4: Lifecycle inference evidence.}
Table~\ref{tab:rq3_lifecycle_ablation} shows that both functional-category evidence and style-change evidence contribute to lifecycle inference. Removing style-change evidence reduces overall F1 from 0.8328 to 0.6935, with the largest effect on \textsc{Changed} records, whose F1 drops from 0.8119 to 0.5769. Without style-change evidence, the model can still see that two records play the same role, but it misses changes in size, hierarchy, spacing, color, or component composition that make the record a meaningful \textsc{Changed} case. Removing functional-category evidence reduces overall F1 to 0.7676 and also weakens \textsc{Changed} recognition. In this variant, visually similar records with different user-facing roles are more likely to be matched, while functionally continuous records with changed appearance are more likely to be split. These results support DesignEvo's cross-artifact matching design: lifecycle inference requires both functional continuity and style-value changes.
The full lifecycle-inference estimate is also stable: precision is 0.8356 [0.8224, 0.8485] and recall is 0.8301 [0.8167, 0.8430] at the 95\% confidence level.

\begin{table}[t]
\centering
\caption{Ablation results for lifecycle-state inference.}
\label{tab:rq3_lifecycle_ablation}
\scriptsize
\setlength{\tabcolsep}{4pt}
\begin{tabular}{lrrr}
\toprule
\textbf{Variant} & \textbf{Prec.} & \textbf{Rec.} & \textbf{F1} \\
\midrule
DesignEvo & 0.8356 & 0.8301 & 0.8328 \\
Without style-change evidence & 0.6958 & 0.6912 & 0.6935 \\
Without functional-category evidence & 0.7702 & 0.7651 & 0.7676 \\
\bottomrule
\end{tabular}
\end{table}

\section{\improve{Discussion}}
\label{sec:discussion}

\subsection{\improve{Implications for Design-System Recovery}}
\label{sec:discussion_evidence}

\improve{The results suggest that design-system recovery benefits from separating geometric localization from semantic interpretation. With GPT-5.5 held constant, candidate-constrained recovery achieves an F1 of 0.8510, compared with 0.4223 for direct localization (Table~\ref{tab:rq1_group_recovery}). By grounding predictions in runtime-derived regions, DesignEvo allows the VLM to focus on functional coherence rather than coordinate generation. However, recovery still depends on candidate coverage and the model's semantic judgment, as reflected in the model-sensitivity results.}

\improve{Repository-level recovery further requires selective aggregation and complementary evidence across versions. RQ3 shows that spatial context helps prevent incorrect family assignments, while frequency weighting distinguishes recurring style conventions from incidental values. For lifecycle inference, functional evidence supports correspondence between records, whereas composition and style changes help distinguish \textsc{Reused} from \textsc{Changed} (Table~\ref{tab:rq3_lifecycle_ablation}). Together with the RQ2 improvements over element-level matching baselines, these findings support recovering and comparing design-system records beyond individual UI elements. Nevertheless, the resulting families and shared styles represent observed implementation patterns, not necessarily intended design rules.}

\subsection{\improve{Implications for Repository-Level Change Review}}
\label{sec:record_use}

\improve{DesignEvo provides traceable relationships between rendered UI regions, repository-level families, shared styles, and their lifecycle states. During change review, a \textsc{Changed} record can direct attention to corresponding occurrences and their structural or stylistic differences, while related family members provide context for whether similar changes appear elsewhere. Likewise, \textsc{Introduced} and \textsc{Removed} identify unmatched records within the analyzed inventories, rather than necessarily indicating application-wide additions or deletions. These relationships offer a structured basis for inspecting UI evolution beyond isolated screenshot or source-code diffs.}

\improve{However, observed differences do not automatically constitute design-system violations. Recurring implementation patterns provide empirical references, whereas conformance judgments require applicable design specifications and evidence of a breach. Intentional variants and revisions may legitimately differ from established patterns. Although the experiments evaluate recovery and lifecycle-inference accuracy, they do not establish whether these records improve human review. Future studies should compare review with and without DesignEvo, measuring affected-pattern identification, false inconsistency reports, and review effort.}

\subsection{\improve{Threats to Validity}}

\improve{\textbf{Construct validity.}
Functional UI groups are an operational choice of granularity, informed by component catalogs and the taxonomy in Section~\ref{sec:method_taxonomy}. They do not capture every individual control state, navigation flow, or behavioral contract. Family boundaries also involve judgment: shared appearance may conceal different purposes, and the same purpose may have several valid implementations. Shared style records introduce further approximation, as grouping similar computed values may combine intentionally distinct styles without recovering their underlying design-token relationships. Responsive layouts and unobserved interaction states may also remain outside the recovered representation. Occurrence references support inspection of these interpretations, but cannot establish intended design rules, design quality, or completeness.}

\improve{\textbf{Internal validity.}
VLM decisions can vary with model version, prompt wording, and image quality. Fixed candidate regions, structured outputs, deterministic inference settings, and rule-based checks constrain this variability, while model-sensitivity analysis examines its effect; none guarantees identical semantic judgments. Errors may also propagate: missed groups affect family assignment, while incorrect family links can alter lifecycle inference. Differences in rendering conditions or observed routes and states may produce apparent additions or removals.}

\improve{\textbf{Annotation and conclusion validity.}
Annotations require judgments about functional boundaries and meaningful changes. Independent review, partial re-annotation, and adjudication reduce disagreement but cannot establish that all plausible decompositions are represented. Lifecycle labels are imbalanced: \textsc{Reused} accounts for 71.01\% of labels, whereas \textsc{Ambiguous} accounts for only 0.12\%. Per-state results complement overall accuracy, but the rarity of ambiguous cases limits conclusions about uncertainty handling. Repeated versions of the same page are also related observations and should not be treated as independent applications.}

\improve{\textbf{External validity.}
The benchmark excludes versions that cannot be built or rendered and focuses on routes with observable UI changes. Results therefore characterize the selected, renderable histories rather than all frontend commits. Applications with sparse component reuse, substantial personalization, canvas-based rendering, or different responsive and interaction states may require additional evidence. Projects maintaining synchronized design-system inventories may also benefit less from recovery. Further evaluation across these settings and on downstream review tasks is needed to establish broader applicability.}

\section{Conclusion} 

This paper presented DesignEvo, an approach for recovering and tracking implemented design-system records across frontend repositories and versions. By combining program-derived candidate regions with VLM-based functional judgments, DesignEvo recovers functional UI groups, aggregates recurring patterns and shared styles into repository-level records, and infers their lifecycle states across commits. Evaluated on 10 open-source repositories, DesignEvo achieves an F1 score of 0.85 for group recovery and an overall lifecycle-state accuracy of 0.84, compared with 0.71 for the strongest lifecycle baseline. Ablation results further demonstrate the contributions of candidate constraints, functional categories, and style evidence. The resulting records provide structured, traceable evidence of how interface patterns are reused and evolve, supporting downstream UI change auditing, design-system drift analysis, and frontend maintenance.

\bibliographystyle{ACM-Reference-Format}
\bibliography{refer}

\appendix


\section{Prompt Templates}
\label{app:prompts}

\begin{promptbox}
{Prompt A:functional UI group recovery}
{box:block}
\begin{lstlisting}[style=promptstyle]
{ 
  "system_message": "You are a functional UI group annotator. Return strict JSON only. You may only select existing candidate_ids. Do not invent or edit bbox coordinates.",
  "user_prompt": {
    "task": "Select and classify functional UI groups from program-generated candidate boxes.",
    "goal": "Identify visible, functionally coherent UI groups on the webpage. The model should decide which candidates should be kept, removed, merged, and assigned to the most appropriate taxonomy label.",
    "page_context": {
      "repo_id": "{repo_id}",
      "page_id": "{page_id}",
      "commit": "{commit_hash}",
      "image_size": {
        "width": "{image_width}",
        "height": "{image_height}"
      }
    },
    "important_principles": [
      "Use functional role and user-facing purpose as the primary basis for judgment.",
      "Use visual grouping, layout boundary, text content, DOM source, interaction target, and candidate evidence as supporting information.",
      "Only output candidate_ids from the provided candidate list.",
      "Do not output new bbox coordinates. The final bbox will be fixed by the program using the selected candidate boxes.",
      "If several candidates together represent one functional UI group, merge them by listing all corresponding candidate_ids in one output group.",
      "If a candidate is only a decorative fragment, an ordinary duplicated parent/child box, or an incomplete part of a larger coherent block, omit it.",
      "Keep meaningful second-level functional UI groups when they have independent function, such as Search, Breadcrumb, Form, Data Table, Visualization, Card, or Action Bar.",
      "When a large region contains several independent functional blocks, prefer preserving the independent blocks rather than only selecting the large container.",
      "When a large region itself has a coherent functional role, such as a full sidebar, full header, complete form, complete table, or complete card panel, keep it as one functional UI group.",
      "Choose the closest taxonomy label from the provided taxonomy definitions. Do not create new taxonomy labels."
    ],
    "preprocessing_rules_used_to_generate_candidates": [
      "Candidates may be generated by grouping visible elements that share the same inferred interactive parent.",
      "Candidates may be generated by grouping visible elements that navigate to the same target href.",
      "Candidates may also be generated from visible DOM containers that have clear functional roles, such as cards, tables, forms, charts, lists, breadcrumbs, search areas, headers, sidebars, and footers."
    ],
    "taxonomy_definitions": [
      {
        "taxonomy_l1": "Global Elements",
        "taxonomy_l2": "Header",
        "definition": "A page-level or site-level top global area, including navigation bars, app bars, top bars, logo areas, account entries, or global actions."
      },
      {
        "taxonomy_l1": "Global Elements",
        "taxonomy_l2": "Footer",
        "definition": "A page-level or site-level bottom global area, usually containing auxiliary navigation, copyright information, link groups, or support information."
      },
      {
        "taxonomy_l1": "Navigation",
        "taxonomy_l2": "Breadcrumb",
        "definition": "A path-style navigation block showing the current page hierarchy or location, often composed of multiple hierarchical items."
      },
      {
        "taxonomy_l1": "Navigation",
        "taxonomy_l2": "Search",
        "definition": "A search module or search area used to query, filter, or locate page content. It may include an input field, icon, button, or filter controls."
      },
      {
        "taxonomy_l1": "Navigation",
        "taxonomy_l2": "Sidebar",
        "definition": "A side navigation or utility area, usually containing multiple navigation entries, menu groups, account entries, settings, notifications, or status-related shortcuts."
      },
      {
        "taxonomy_l1": "Action",
        "taxonomy_l2": "Action Bar",
        "definition": "A command region that groups actions for the same page, section, item, or selected set, such as edit, delete, share, filter, sort, save, export, or batch operations."
      },
      {
        "taxonomy_l1": "Input",
        "taxonomy_l2": "Input",
        "definition": "A field-centered input block that groups one input goal with supporting suggestions, options, or filtering behavior, such as a text field, text area, autocomplete field, combobox input, or filter input."
      },
      {
        "taxonomy_l1": "Input",
        "taxonomy_l2": "Picker/Dropdown Selector",
        "definition": "A selection block that groups options for choosing a value, view, mode, date, or state."
      },
      {
        "taxonomy_l1": "Input",
        "taxonomy_l2": "Form",
        "definition": "A task-oriented input region that groups multiple fields, labels, helper text, validation messages, and actions for submission, filtering, or configuration."
      },
      {
        "taxonomy_l1": "Data Display",
        "taxonomy_l2": "Card",
        "definition": "A bounded content container that groups related title, text, media, data, actions, or status information into one coherent visual unit."
      },
      {
        "taxonomy_l1": "Data Display",
        "taxonomy_l2": "Stats Panel",
        "definition": "A summary region that groups multiple key values or metrics for quick comparison or monitoring."
      },
      {
        "taxonomy_l1": "Data Display",
        "taxonomy_l2": "Visualization",
        "definition": "A data presentation block that groups visual marks, labels, legends, or controls for understanding values or patterns, such as charts, graphs, or maps."
      },
      {
        "taxonomy_l1": "Data Display",
        "taxonomy_l2": "Timeline",
        "definition": "A block that organizes events, activities, stages, updates, or logs in temporal order."
      },
      {
        "taxonomy_l1": "Data Display",
        "taxonomy_l2": "Grid",
        "definition": "A grid-based content display block in which multiple repeated content units are arranged in a two-dimensional structure, such as card grids, metric grids, or image grids."
      },
      {
        "taxonomy_l1": "Data Display",
        "taxonomy_l2": "Code Snippet",
        "definition": "A code display block, often including code content, syntax highlighting, copy actions, labels, or language indicators."
      },
      {
        "taxonomy_l1": "Data Display",
        "taxonomy_l2": "Data Table",
        "definition": "A structured data display block with rows, columns, table headers, cells, statuses, actions, or pagination-related table controls."
      },
      {
        "taxonomy_l1": "Data Display",
        "taxonomy_l2": "List",
        "definition": "A list-style display block composed of multiple list items, menu items, action items, records, or repeated textual entries. Page-level navigation should not be classified as List when Sidebar or Header is more appropriate."
      },
      {
        "taxonomy_l1": "Data Display",
        "taxonomy_l2": "Hero",
        "definition": "A main visual or introductory block, usually containing a headline, description, image, call-to-action, or major promotional content."
      },
      {
        "taxonomy_l1": "Data Display",
        "taxonomy_l2": "Content Block",
        "definition": "A general content emphasis or summary block, usually composed of title, text, image, callout, description, or supporting action, when no more specific taxonomy label fits better."
      },
      {
        "taxonomy_l1": "Status",
        "taxonomy_l2": "Message",
        "definition": "A feedback region that groups a status message with related context, severity, or dismissal controls, including alert banners, inline messages, snackbars, toasts, warnings, or success messages."
      },
      {
        "taxonomy_l1": "Status",
        "taxonomy_l2": "Empty State",
        "definition": "A block shown when an area has no content, no result, no data, or no available item, usually with explanatory text or an action suggestion."
      },
      {
        "taxonomy_l1": "Status",
        "taxonomy_l2": "Loader",
        "definition": "A loading feedback region that indicates pending content or an ongoing operation, such as a loading spinner or skeleton loader."
      },
      {
        "taxonomy_l1": "Status",
        "taxonomy_l2": "Progress",
        "definition": "A progress feedback region that shows completion state across a task, process, or sequence."
      },
      {
        "taxonomy_l1": "Overlay",
        "taxonomy_l2": "Dialog/Modal",
        "definition": "An overlay region that groups temporary content and related actions above the current page, such as a dialog, modal, bottom sheet, or popup."
      },
      {
        "taxonomy_l1": "Image Types",
        "taxonomy_l2": "Logo/Ad Block",
        "definition": "An image-based identity or promotional block that presents brand marks, campaign content, sponsorship labels, links, or actions as one recognizable visual region."
      },
      {
        "taxonomy_l1": "Third-party Embedded",
        "taxonomy_l2": "Embedded Widget",
        "definition": "An externally embedded or third-party module, such as an iframe, widget, ticker, social component, or externally hosted interactive component."
      }
    ],
    "candidate_input_schema": {
      "candidate_id": "Unique id of the program-generated candidate box.",
      "source": "How the candidate was generated, such as DOM container, shared interactive parent, same href target, or functional container detection.",
      "bbox": "Program-generated candidate box. The model must not edit it.",
      "text": "Visible text extracted from the candidate region.",
      "href": "Navigation target if available.",
      "suggested_taxonomy": "A weak program suggestion, used only as auxiliary evidence."
    },
    "output_schema": {
      "groups": [
        {
          "label": "A short human-readable English name for the functional UI group.",
          "taxonomy_l1": "One first-level taxonomy name from taxonomy_definitions.",
          "taxonomy_l2": "One second-level taxonomy label from taxonomy_definitions.",
          "candidate_ids": [
            "candidate ids selected or merged into this group"
          ],
          "confidence": "high | medium | low",
          "reason": "Briefly explain why the selected candidate(s) form this functional UI group."
        }
      ]
    },
    "candidates": "{candidate_list}"
  }
}
\end{lstlisting}
\end{promptbox}

\begin{promptbox}
{Prompt B: UI group family assignment}
{box:semantic}
\begin{lstlisting}[style=promptstyle]
{
  "stage": "ui_group_family_assignment",
  "prompt_template": {
    "task": "Decide whether two functional UI groups from the same webpage should be variants of the same-version UI group family.",
    "page_key": "...",
    "ablation_variant": "full",
    "decision_policy": [
      "Return same_family=true only when the two groups have the same primary user-facing purpose and look like repeated variants of one UI group family.",
      "If functional category or user-facing purpose differs, same_family must be false even if style and size are similar.",
      "If size of bbox differs, same_family should be false.",
      "If style is clearly different, same_family should be false unless it is an intentional visual variant of the same repeated UI group family.",
      "Examples of same families: KPI cards in one dashboard row, pricing plan cards, repeated alert panels, repeated profile information cards.",
      "Examples of different families: Header vs Sidebar, Hero vs Card, Visualization card vs KPI card, Form/control vs data card.",
      "If a group was mislabeled as Form but its reason says it is a KPI/metric card, it may belong to a visually matching Card family.",
      "Return JSON only."
    ],
    "output_shape": {
      "pairs": [
        {
          "pair_id": "p001",
          "same_family": true,
          "reason": "both are repeated KPI summary cards with compatible size/style"
        }
      ]
    },
    "pairs": [
      "... candidate group pairs ..."
    ]
  }
}
\end{lstlisting}
\end{promptbox}

\begin{promptbox}
{Prompt C: Lifecycle state detection}
{box:lifecycle}
\begin{lstlisting}[style=promptstyle]
{
  "stage": "lifecycle_inference",
  "prompt_template": {
    "task": "Infer lifecycle states for functional UI group family records across adjacent commits.",
    "page": "...",
    "states": [
      "Reused",
      "Changed",
      "Introduced",
      "Removed",
      "Ambiguous"
    ],
    "rules": [
      "Match by role/function and taxonomy",
      "Reused=same repository-level record without meaningful visual, structural, or style value change.",
      "Changed=same repository-level record with meaningful layout, size, style value, content arrangement, UI element composition, or convention-level change.",
      "Introduced=current record has no previous counterpart.",
      "Removed=previous record has no current counterpart.",
      "Ambiguous=noisy, unreliable, or insufficiently supported correspondence.",
      "No fixed numeric thresholds in reasons. Strict JSON only."
    ],
    "prev": [
      "... previous commit records ..."
    ],
    "cur": [
      "... current commit records ..."
    ],
    "pairs": [
      "... candidate cross-commit matches ..."
    ],
    "schema": {
      "matches": [
        {
          "prev_record_id": "R1",
          "current_record_id": "R1",
          "state": "Reused",
          "confidence": 0.9,
          "reason": "..."
        }
      ],
      "introduced": [
        {
          "current_record_id": "R9",
          "state": "Introduced",
          "confidence": 0.8,
          "reason": "..."
        }
      ],
      "removed": [
        {
          "prev_record_id": "R8",
          "state": "Removed",
          "confidence": 0.8,
          "reason": "..."
        }
      ]
    }
  }
}
\end{lstlisting}
\end{promptbox}


\section{Functional UI Group Taxonomy}
\label{app:functional_ui_group_taxonomy}
Table~\ref{app:functional_group_taxonomy} lists the functional UI group taxonomy used in DesignEvo. Each label represents a functionally coherent page region: its elements are grouped because they jointly serve the same user-facing purpose, such as navigation, input, data display, status feedback, overlay interaction, identity, promotion, or third-party embedding. The taxonomy therefore preserves functional UI groups at the region level, while recording source component names as evidence of the design-system component terms merged into each group label.

\begin{table*}[t]
\centering
\caption{Functional UI group taxonomy used by DesignEvo.}
\label{app:functional_group_taxonomy}
\scriptsize
\setlength{\tabcolsep}{3pt}
\renewcommand{\arraystretch}{1.08}
\begin{tabular}{p{0.15\textwidth}p{0.17\textwidth}p{0.25\textwidth}p{0.35\textwidth}}
\toprule
\textbf{Primary Taxonomy} & \textbf{Secondary Taxonomy} & \textbf{Source Components} & \textbf{Description} \\
\midrule
Global Elements & Header & Header; App Bar; Navigation Bar & Top page or site region that groups global navigation, branding, and page-level controls. \\
Global Elements & Footer & Footer & Bottom page or site region that groups auxiliary navigation, copyright text, and related links. \\
Navigation & Breadcrumb & Breadcrumb & Path navigation block that groups ordered hierarchy items showing the user's current location. \\
Navigation & Search & Search & Search block that groups a query field with related controls such as icons, filters, or suggestions. \\
Navigation & Sidebar & Sidebar; Utility Bar; Global Bar & Side region that groups navigation entries and related utility entries such as account, settings, or notifications. \\
Action & Action Bar & Action Bar & Command region that groups actions for the same page, section, item, or selected set. \\
Input & Input & Autocomplete; Combobox; Filter Input; Text Field; Text Area & Field-centered input block that groups one input goal with supporting suggestions, options, or filtering behavior. \\
Input & Picker/Dropdown Selector & Date Picker; Select; Dropdown; Segmented Control & Selection block that groups options for choosing a value, view, mode, or state. \\
Input & Form & Form; Filter Form & Task-oriented input region that groups multiple fields, labels, helper text, and actions for submission, filtering, or configuration. \\
Data Display & Card & Cards; Tile; Accordion; Collapsible Section & Self-contained content or task block that groups related text, media, metadata, and actions. \\
Data Display & Stats Panel & Stats Panel; Metrics Panel; KPI Panel & Summary region that groups multiple key values or metrics for quick comparison or monitoring. \\
Data Display & Visualization & Chart; Graph; Map & Data presentation block that groups visual marks, labels, legends, or controls for understanding values or patterns. \\
Data Display & Timeline & Timeline; Activity Component; Activity Feed & Time-ordered region that groups events, activities, stages, updates, or logs. \\
Data Display & Grid & Card Grid; Grid Layout & Repeated-content display region that groups similar items in a two-dimensional arrangement. \\
Data Display & Code Snippet & Code Snippet & Code presentation block that groups code with related labels, language indicators, or copy actions. \\
Data Display & Data Table & Data Table & Structured data block that groups rows, columns, headers, and related status or control areas. \\
Data Display & List & List; Menus; Action List & Repeated-item region that groups similar entries, including menu-like or action-list entries when they form one coherent block. \\
Data Display & Hero & Hero Units; Hero Block; Hero Banner; Carousel & Prominent introductory region that groups headline content, supporting text, media, and primary actions. \\
Data Display & Content Block & Callouts; Touts; Summaries & Emphasis or summary region that groups related title, body text, media, or action content. \\
Status & Message & Alert Banner; Inline Message; Snackbar; Warning Message; Toast Notification & Feedback region that groups a status message with related context, severity, or dismissal controls. \\
Status & Empty State & Empty State & Placeholder region that explains the absence of content, results, data, or items and may include a recovery action. \\
Status & Loader & Loading Spinner; Skeleton Loader & Loading feedback region that indicates pending content or an ongoing operation. \\
Status & Progress & Step Indicator; Progress Bar & Progress feedback region that shows completion state across a task, process, or sequence. \\
Overlay & Dialog/Modal & Dialog; Modal; Bottom Sheet; Popup & Overlay region that groups temporary content and related actions above the current page. \\
Image Types & Logo/Ad Block & Logos; Brand Logo; Advertising; Ads; Ad Formats; Promo; Marketing Banner & Image-based identity or promotional block that presents brand marks, campaign content, sponsorship labels, links, or actions as one recognizable visual region. \\
Third-party Embedded & Embedded Widget & Widgets; Iframes; Stock Tickers; Social Buttons; Third-party Components & Embedded external module that appears as one recognizable functional region within the page. \\
\bottomrule
\end{tabular}
\end{table*}

\section{Functional UI Group Distribution}
\label{app:functional_ui_group_distribution}
Table~\ref{tab:functional_ui_group_distribution} reports the distribution of annotated functional UI groups in the evaluation dataset at the primary and secondary taxonomy levels. Each count refers to one annotated functional UI group. Data Display is the largest primary category, accounting for 58.25\% of all annotated groups, followed by Navigation at 24.15\%. At the secondary level, Card is the most frequent type. The dataset also includes global page regions, navigation regions, input regions, status regions, and command regions, reflecting the recurring functional UI groups observed in the evaluation dataset.

This distribution should be interpreted as dataset coverage rather than taxonomy coverage. The taxonomy in Table~\ref{app:functional_group_taxonomy} defines the functional UI group label space used by DesignEvo, whereas Table~\ref{tab:functional_ui_group_distribution} reports only the group types observed in the sampled mobile screens. Therefore, some taxonomy categories are absent or sparse in this table because they occur infrequently in the evaluation dataset, not because they are outside the functional scope of DesignEvo.

\begin{table*}[t]
\centering
\caption{Distribution of primary and secondary functional UI group types. The evaluation dataset contains 3,478 annotated functional UI groups.}
\label{tab:functional_ui_group_distribution}
\footnotesize
\setlength{\tabcolsep}{7pt}
\renewcommand{\arraystretch}{1.06}
\begin{tabular}{llrr}
\toprule
\textbf{Primary Taxonomy} &
\textbf{Secondary Taxonomy} &
\textbf{Groups} &
\textbf{Percentage} \\
\midrule

\multirow{8}{*}{\shortstack[l]{Data Display\\2,026 (58.25\%)}}
& Card                  & 1,142 & 32.83\% \\
& Stats Panel           & 292   & 8.40\%  \\
& Data Table            & 209   & 6.01\%  \\
& Visualization         & 166   & 4.77\%  \\
& List                  & 128   & 3.68\%  \\
& Content Block         & 60    & 1.73\%  \\
& Timeline              & 18    & 0.52\%  \\
& Hero                  & 11    & 0.32\%  \\
\cmidrule(lr){1-4}

\multirow{3}{*}{\shortstack[l]{Navigation\\840 (24.15\%)}}
& Sidebar               & 305   & 8.77\%  \\
& Breadcrumb            & 297   & 8.54\%  \\
& Search                & 238   & 6.84\%  \\
\cmidrule(lr){1-4}

\multirow{2}{*}{\shortstack[l]{Global Elements\\326 (9.37\%)}}
& Header                & 286   & 8.22\%  \\
& Footer                & 40    & 1.15\%  \\
\cmidrule(lr){1-4}

\multirow{2}{*}{\shortstack[l]{Input\\251 (7.22\%)}}
& Form                  & 150   & 4.31\%  \\
& Picker/Dropdown Selector
                        & 101   & 2.90\%  \\
\cmidrule(lr){1-4}

\shortstack[l]{Status\\32 (0.92\%)}
& Message               & 32    & 0.92\%  \\
\cmidrule(lr){1-4}

\shortstack[l]{Action\\3 (0.09\%)}
& Action Bar            & 3     & 0.09\%  \\
\midrule

\multicolumn{2}{l}{\textbf{Total}}
& \textbf{3,478}
& \textbf{100.00\%} \\
\bottomrule
\end{tabular}
\end{table*}

\end{document}